\documentclass[lettersize,journal]{IEEEtran}
\usepackage{amsmath,amsfonts, amssymb, amsthm}
\usepackage{algorithmic}
\usepackage{algorithm}
\usepackage{array}
\usepackage[caption=false,font=normalsize,labelfont=sf,textfont=sf]{subfig}
\usepackage{textcomp}
\usepackage[T2A,T1]{fontenc}
\newcommand{\LetterSha}{\text{\fontencoding{T2A}\selectfont Ш}}
\usepackage{stfloats}
\usepackage{url}
\usepackage{verbatim}
\usepackage{graphicx}
\usepackage{cite}
\usepackage{siunitx}
\DeclareSIUnit\angstrom{\text {Å}}
\usepackage{mathrsfs}
\usepackage{hyperref}
\begin{document}

\title{Super-Resolution Structured-Illumination X-Ray Microscopy based on Fourier Decomposition}

\author{Stefan Schwaiger, Lennart Forster, Martin Dierolf, Franz Pfeiffer and Benedikt Günther
\thanks{All authors are with the Chair of Biomedical Physics, Department of Physics, TUM School of Natural Sciences, and Munich Institute of Biomedical Engineering, Technical University of Munich, 85748 Garching, Germany.\\
\indent Franz Pfeiffer is additionally with the TUM Institute for Advanced Study, Technical University of Munich, 85748 Garching, Germany.}
}

\maketitle

\begin{abstract}
X-ray microscopy has become an important tool for non-destructive testing, e.g., in battery research. However, imaging a cm-scale battery cell at the desired (sub-)micrometer resolution has been challenging.
State-of-the-art X-ray microscopy techniques with a suited field-of-view provide (sub-) 10\,\textmu m resolution, typically limited by the detector point-spread function and the (effective) detector pixel size.
This work presents a super-resolution X-ray microscopy approach overcoming both limitations.
It requires a structured X-ray illumination to encode high-frequency sample information that is natively unresolved within the resolved region of support.
A mathematical framework is developed that decodes this information and generates a super-resolved image from multiple acquisitions with different phase shifts of the structured X-ray illumination.
The presence of this encoded high-frequency information is first experimentally demonstrated, followed by quantification and validation using a resolution test chart.
A resolution improvement by a factor of 2.2 is shown.
Finally, we extend the proposed super-resolution technique to X-ray microtomography.
Since the image acquisition scheme is inherently multimodal, phase-contrast and dark-field X-ray images can be computed additionally.
These results showcase the direct impact of the proposed technique across both non-destructive testing and biomedical imaging, alleviating pixel-size limitations in detectors and sample-size restrictions.
\end{abstract}


\section{Introduction}
\IEEEPARstart{X}{-ray} transmission microscopy is a widely researched technique due to its high spatial resolution and its ability to penetrate matter opaque to visible light.
This is particularly relevant for non-destructive testing in fields such as additive manufacturing \cite{thompson_x-ray_2016,rathore_-depth_2025} and battery engineering \cite{villarraga-gomez_assessing_2022}, as well as in biomedical research, including virtual histology \cite{schaeper_3d_2025,busse_three-dimensional_2018} and high-resolution imaging of small animals \cite{preuss_functional_2025}.
Achieving resolutions comparable to or exceeding those of visible light microscopy has been the focus of a large part of the body of research.
Early efforts, such as those by Sievert et al.\ \cite{sievert_two_1936}, proposed a methodology involving X-ray projection onto photographic film and subsequent magnification in a visible light microscope.
First advances superseding this approach were made possible by better electron beam focusing optics, which enabled precise control of the electron spot size, thus reducing source blur.
This was first realized by Cosslett et al.\ \cite{cosslett_x-ray_1952} in a transmission target configuration. \newline
The refinement of X-ray transmission microscopy approaches towards their contemporary, well-established form was initiated by the development of X-ray optics, e.g., the KB-mirror \cite{kirkpatrick_formation_1948}, the compound refractive lens \cite{snigirev_compound_1996}, and diffractive optics such as the Fresnel zone plate.
Apart from typical geometric magnification microscopy, X-ray optics are generally applied in two distinct experimental geometries: full-field transmission X-ray microscopy (TXM) and scanning transmission X-ray microscopy (STXM).
In a TXM setup, the sample is positioned at the focus of a condenser optic. The diverging beam is then refocused by an objective onto a 2D pixelated detector.
In an STXM setup, the detector records the intensity transmitted through the sample at the focal point position.
To image a larger section of the sample, it is scanned at multiple positions, with modern setups achieving resolutions of up to $\qty{30}{\nano\meter}$ \cite{feggeler_scanning_2023}.
Reaching such resolutions requires focal spot sizes in the order of $\qty{10}{\nano\meter}$, which imposes stringent constraints on sample dimensions and acquisition time.
Contrary to this limitation, \cite{gunther_full-field_2019} proposed a hybrid approach that merges STXM and Talbot array illumination.
Rather than using an X-ray focusing optic to illuminate a single spot on the sample, the Talbot effect \cite{talbot_lxxvi_1836,rayleigh_xxv_1881} is employed to project an array of individual beamlets onto the sample.
Scanning this array over a single period enables parallelization of the STXM scanning approach, thereby drastically reducing acquisition time.
This method has been shown to achieve a resolution of several $\qty{100}{\nano\meter}$ \cite{gunther_full-field_2019}. \newline
In this publication, we propose structured-illumination super-resolution X-ray microscopy inspired by a well-established visible light microscopy technique \cite{heintzmann_laterally_1999,gustafsson_surpassing_2000}.
Contrary to \cite{gunther_full-field_2019}, who treated the illumination as a collection of discrete focal spots, our method describes the illumination in terms of its Fourier spectral composition.
The intensity distribution of a periodically structured illumination, generated with a grating by means of the Talbot effect \cite{talbot_lxxvi_1836,rayleigh_xxv_1881}, is modulated by sample interaction and acquired on a 2D detector.
The Fourier transform of the intensity field at the detector plane prior to detection is described as a convolution of the sample's Fourier spectrum and a reciprocal Dirac comb, with frequencies determined by the pattern's fundamental frequency and its harmonics.
Consequently, the intensity field is a superposition of replicas of the sample spectrum, with each replica shifted in Fourier space to the position of the peak that gives rise to it.
Hence, around each peak in the reciprocal Dirac comb of the pattern, the same sample information populates the frequency landscape, except for a linear factor arising from the amplitude and phase of the peak.
This frequency-space superposition is formed independently of detection, meaning it is present despite the optical transfer function (OTF) or the discretization introduced during detection.
Therefore, accessing the additional information encoded in the superposition enables the reconstruction of a super-resolved image with an extended pass-band of $\mathbf{k}_\mathrm{n} + N\mathbf{k}_\mathrm{p}$, where $\mathbf{k}_\mathrm{n}$ is the native pass-band, $N$ is the highest reconstructed order, and $\mathbf{k}_\mathrm{p}$ is the pattern frequency.
To this end, a series of projections is acquired with varying positions of the illumination pattern.
This defines a linear system of equations for which a synthesis matrix can be constructed, which assembles each projection from a basis described by the set of sample replicas.
The solution of this set of equations produces a set of Fourier space images, each containing only the information contributed by a single replica. \newline
After shifting the separated components to the origin in Fourier space and accounting for a linear factor, the components can be combined using a Wiener filter.
This can be used to retrieve a sample image containing information surpassing the limits imposed by the OTF and the Nyquist frequency, provided the reconstruction sampling satisfies the Nyquist criterion for the super-resolved pass-band.

\section{Theory}
\subsection{Encoding Additional High-Frequency Information}\label{sec:additionalinfo}
\noindent Attenuation-based imaging contrast is commonly described by the Beer--Lambert law
\begin{align}
\label{eq:lambertbeer}
\begin{split}
    I(x,y) =& \int_0^{E_\mathrm{max}} I_0(E,x,y)\\
    & \cdot \;\exp\left[-\int_0^d \mu(E, \eta, x, y)\mathrm{d}\eta\right]\mathrm{d}E,
    \end{split}
\end{align}
where $I(x,y)$ is the intensity function after interaction with a sample, $I_0(E,x,y)$ is the incident spectral intensity distribution, and $\mu(E, \eta, x, y)$ denotes the linear attenuation coefficient as a function of photon energy $E$, depth within the sample $\eta$, and transverse position  $(x,y)$.
The variable $d$ represents the total sample thickness.
For the scope of the proposed proof-of-principle, it is sufficient to assume monochromaticity and abbreviate all sample-induced attenuation interaction into the function $S(x,y)$, yielding
\begin{equation}
    I(x,y) = \left[I_0(x,y) \cdot S(x,y)\right] \otimes h(x,y).
\end{equation}
where the additional convolution with the PSF $h(x,y)$ is induced by detection.
Given that our method aims to increase resolution by exploiting the effects of periodic illumination on the spatial frequency content of an acquisition, it is natural to consider the Fourier-domain representation of this model.
The Fourier transform of this product gives
\begin{align}
\label{eq:fundamentalsimicfreq}
\begin{split}
    \mathscr{F}\{I(x,y)\}(u,v) =& \big[ \mathscr{F}\{ I_0(x,y) \}(u,v)\\
    & \otimes \mathscr{F}\{ S(x,y) \}(u,v) \big]\cdot O(u,v),
\end{split}
\end{align}
where $\mathscr{F}$ denotes the Fourier transform operator, $\otimes$ denotes the convolution arising from the convolution theorem, and $O(u,v) = \mathscr{F}\{h(x,y)\}(u,v)$ is the OTF.
Since the illumination is presumed infinitely periodic, its spatial substructure may be described by five possible Bravais lattices. 
Following the rationale of \cite{schropp_twodimensional_2014}, only the square and hexagonal lattices are well suited for our purpose of Fourier-space analysis, as their basis vector magnitude is equal along both axes, thus populating the frequency space isotropically.
Given that \cite{gustschin_high-resolution_2021} have demonstrated the applicability of high-visibility two-dimensional square phase gratings, the illumination function $I_0(x,y)$ is henceforth assumed as an infinitely periodic square lattice:
\begin{align}
\begin{split}
\label{eq:realspaceintensity}
    I_0(x,y) =& \bar{I}+a\left[ \mathrm{rect}\left((x-x_0)\, \frac{d}{wp}, (y-y_0)\, \frac{d}{wp}\right) - \frac{w}{d} \right]\\
    &\otimes K(x,y) \otimes \LetterSha_\mathrm{p, p},
\end{split}
\end{align}
where $\mathrm{rect}$ is the rectangular function and $\LetterSha$ is the Dirac comb.
The arbitrary shift vector $(x_0, y_0)$ originates from the possible asymmetry of the illumination pattern with respect to the coordinate origin.
The convolution kernel $K$ describes the blurring of the idealized illumination structure due to grating imperfections or partial coherence.
The parameter $p$ denotes the pattern period, and $w/d$ is the opening ratio of the illumination.
The factor $a$ is the amplitude modulation, while $\bar{I}$ is the mean intensity.
Under Fourier transformation, the convolution is expressed as a multiplication, resulting in
\begin{align}
\label{eq:illspectrum}
\begin{split}
   \mathscr{F}\{ I_0(x,y) \}(u,v) = & I_\mathrm{a}\,\mathrm{e}^{-2\pi i(ux_0 + vy_0)}\left(p\, \frac{w}{d}\right)^2 \\
   &\cdot\mathrm{sinc}_\pi\left( p\,\frac{w}{d}\,u \right) \mathrm{sinc}_\pi\left( p\, \frac{w}{d}\,v \right)\\
   &\cdot\mathscr{F}\{K(x,y)\}(u,v) \cdot \LetterSha_\mathrm{q,q} \\
   \equiv& A(u,v)\mathrm{e}^{i\varphi_0(u,v)}\cdot \LetterSha_\mathrm{q,q},
\end{split}
\end{align}
where $\mathrm{sinc}_\pi=\frac{sin(\pi t)}{\pi t}$ denotes the normalised $\mathrm{sinc}$ function arising from the Fourier transform of the rectangular function.
The reciprocal Dirac comb $\LetterSha_\mathrm{q, q}$ denotes the Fourier transform of the Dirac comb $\LetterSha_\mathrm{p,p}$.
The frequency $q=\frac{1}{p}$ is the fundamental spatial frequency of the illumination pattern.
The modulated intensity
\begin{equation}
    I_\mathrm{a} = 
    \begin{cases}
        \bar{I} & u,v=0,\\
        a\bar{I} & \text{otherwise}
    \end{cases}
\end{equation}
captures the case of pattern modulation $a$ unequal to unity.
The phase ramp $\exp(-2\pi i(ux_0 + vy_0))\equiv\exp(i\varphi_0(u,v))$ originates from a possible asymmetry of the illumination pattern with respect to the origin.
The prefactor $A(u,v)$ captures the amplitude behavior, including the effect of the blurring kernel $K(x,y)$ whose Fourier transform becomes a multiplicative factor.
Substituting Eq.\ \eqref{eq:illspectrum} into Eq.\ \eqref{eq:fundamentalsimicfreq} yields
\begin{align}
\label{eq:superpos}
\begin{split}
    \mathscr{F}\{ I(x,y) \}(u,v)  =& \bigg[\left( A(u,v)\mathrm{e}^{i\varphi_0(u,v)}\cdot \LetterSha_\mathrm{q, q} \right)\\
    & \otimes \mathscr{F}\{ S(x,y) \}(u,v)\bigg]\cdot O(u,v) \\
     =& \sum_\mathrm{n,m=-\infty}^\infty A_\mathrm{(n,m)}\mathrm{e}^{i\varphi_\mathrm{0,(n,m)}}\;O(u,v)\\\
     &\cdot\mathscr{F}\{ S(x,y) \}(u-nq, v-mq).
\end{split}
\end{align}
So the Fourier transform of the intensity after the sample is a superposition of shifted sample-signal replicas, each centered at the respective (n,m)-th harmonic of the grating frequency.
Each replic is multiplied with the corresponding prefactor $A_\mathrm{(n,m)}$ and the constant phase factors $\exp(i\varphi_\mathrm{0, (n,m)})$.
Upon detection, the entire superposition is multiplied by the OTF $O(u,v)$.
Notably, Eq.\ \eqref {eq:superpos} in principle describes a superposition of infinitely many such replicas, provided that the coefficients $A_\mathrm{(n,m)}$ remain non-zero.

\subsection{Recovering Additional Information}\label{sec:recoveringadditionalinformation}
\noindent Let $I_\mathrm{i}(x,y)$ be the intensity distribution, shifted by $(x_\mathrm{i}, y_\mathrm{i})$, after it traversed the sample.
Using Eq.\ \eqref{eq:superpos}, the Fourier-domain representation can be written as
\begin{align}
\begin{split}
    \mathscr{F}\{I_\mathrm{i}(x,y)\}(u,v)  =& \big[\mathscr{F}\{I_\mathrm{0}(x - x_\mathrm{i}, y - y_\mathrm{i}) \}\\
    &\otimes \mathscr{F}\{ S(x,y) \}(u,v)\big]\cdot O(u,v) \\
     =& \sum_\mathrm{n,m=-\infty}^\infty \left. \mathrm{e}^{-2\pi i (ux_i+vy_i)}\right|_\mathrm{(u,v)=(nq, mq)}  \\
    & \cdot A_\mathrm{(n,m)}\mathrm{e}^{i\varphi_\mathrm{0,(n,m)}}\; O(u,v)\\
    &\cdot\mathscr{F}\{ S(x,y) \}(u-nq, v-mq)
\end{split}
\end{align}
where the transverse pattern movement introduces a phase ramp in the corresponding dimension in Fourier space, which becomes discretized into a constant phase factor for each peak of the Dirac comb.
Under convolution with the sample spectrum, each shifted replica of the sample spectrum inherits the corresponding constant phase factor.
Due to the convolutional kernel $K(x,y)$, we assume that only a finite number of contributions to the sum have amplitudes higher than the noise level.
For $s_\mathrm{u}$ peaks in dimension $u$ and $s_\mathrm{v}$ peaks in dimensions $v$, a total number of $N=s_\mathrm{u}s_\mathrm{v}$ pattern translations with $s_\mathrm{u}$ steps in $u$ and $s_\mathrm{v}$ in $v$ are required to assign a distinct phase value to each peak.
Thus, we can confine the harmonic indices to
\begin{align}
\begin{split}
    m \in \left\{ -\frac{s_\mathrm{u}-1}{2},..., \frac{s_\mathrm{u}-1}{2}\right\},\\
    n \in \left\{ -\frac{s_\mathrm{v}-1}{2},..., \frac{s_\mathrm{v}-1}{2}\right\},
\end{split}
\end{align}
which identify each harmonic displacement in the $u$- and $v$- directions, respectively.
Similarly, the indices
\begin{equation}
    \alpha \in \{0, ..., s_\mathrm{u}-1\}, \qquad \gamma \in \{0, ..., s_\mathrm{v}-1\}
\end{equation}
identify the associated pattern phase shift introduced during the raster scan.
The phase factors associated with the harmonics $(m,n)$ at the illumination steps $(\alpha, \gamma)$ are described by the 4D array
\begin{equation}\label{eq:separation_matrix_full}
     M_{\alpha m\gamma n} = \exp\left[ i\, (\phi^u_{\alpha\gamma}\, m + \phi^v_{\alpha\gamma}\, n) \right].
\end{equation}
$\phi^\mathrm{u,v}_{\alpha, \gamma}$ are elements of the set of all combinations of phases in $(u,v)$.
Further, defining
\begin{equation}
    i \equiv \gamma+\alpha\cdot s_\mathrm{v}
    \quad\text{and}\quad
    j \equiv n^{\prime}+m^{\prime}\cdot s_\mathrm{u},
\end{equation}
with $m^{\prime}=m+(s_\mathrm{u}-1)/2$ and $n^{\prime}=n+(s_\mathrm{v}-1)/2$, enables the construction of the matrix
\begin{equation}\label{eq:separation_matrix}
    M_\mathrm{ij} = M_\mathrm{\gamma +\alpha\cdot s_\mathrm{v}, n^{\prime} + m^{\prime}\cdot s_\mathrm{u}}
\end{equation}
of size $(s_\mathrm{u}s_\mathrm{v})\times(s_\mathrm{u}s_\mathrm{v})$.
Thus, $N$ image acquisitions in different pattern positions define a linear system \cite{heintzmann_saturated_2003} as
\begin{equation}\label{eq:linearsystem}
    \Tilde{I}_\mathrm{i}(u,v) = M_\mathrm{ij}\, \Tilde{C}_\mathrm{j}(u,v),
\end{equation}
where $\Tilde{I}_\mathrm{i}=\mathscr{F}_\mathrm{x,y}\{I_\mathrm{i}\}$ and $\Tilde{C}_\mathrm{j}=\mathscr{F}_\mathrm{x,y}\{C_\mathrm{j}\}$, and $i\in\{0,...,N-1\}$.
The matrix $M_\mathrm{ij}\in\mathbb{C}^{N\times N}$, with $j\in\{0,..., N-1\}$ indexing the $N$ components contributing to the superposition in Eq.\ \eqref{eq:superpos}, synthesizes the $i$-th image $\Tilde{I}_\mathrm{i}(x,y)$ from its corresponding set of basis components $\Tilde{C}_\mathrm{j}(x,y)$, each of which contains only the information contributed by a single shifted replica of the sample spectrum.
So long as $M_\mathrm{ij}$ is invertible, Eq.\ \eqref{eq:linearsystem} can be solved to reconstruct the basis components $\Tilde{C}_\mathrm{j}(u,v)$ of the form
\begin{align}
\label{eq:componentform}
\begin{split}
    \Tilde{C}_\mathrm{j}(u,v)=&\Tilde{C}_\mathrm{(n,m)}(u,v) \\
    =&A_\mathrm{(n,m)}\mathrm{e}^{i\varphi_{0,\mathrm{(n,m)}}}\; O(u,v)\\
    &\cdot \mathscr{F}\{{S(x,y)}\}(u-nq, v-mq).
\end{split}
\end{align}
Following the component expression in Eq.\ \eqref{eq:componentform}, the parameters $q$, $A_\mathrm{(n,m)}$, and $\varphi_\mathrm{0,(n,m)}$ must be accurately estimated to reliably extract the portion of the sample spectrum embedded in the separated components.
To determine the shift vector $\mathbf{q} = (nq, mq)^\mathrm{T}$, we exploit that all components are frequency-shifted replicas of the sample spectrum.
Consequently, wherever their information overlaps, this offers a reliable basis for the cross-correlation approach first described by \cite{gustafsson_three-dimensional_2008}.
To compute this cross-correlation, the OTF associated with each separated component in Eq.\ \eqref{eq:linearsystem} is removed by defining $\Tilde{D}_\mathrm{(n,m)}(u,v)\equiv\Tilde{C}_\mathrm{(n,m)}(u,v)O^*(u,v)/|O(u,v)|^2$, which is used only in this process of parameter estimation, but not in the final reconstruction.
The shift vector is decomposed into an integer and a sub-integer component.
The integer shift vector component is obtained straightforwardly using cross-correlation as
\begin{equation}
    \mathbf{q}_{\mathrm{int}} = \mathop{\arg\max}\limits_{u,v \in \mathscr{F},\; n,m\in\mathbb{Z}} (\Tilde{D}_{(0,0)}(u,v)\star \Tilde{D}_{(n,m)}(u,v)),
\end{equation}
where $\star$ denotes the correlation operator.
After shifting each component by this vector, only a subpixel displacement remains.
To accurately estimate this residual vector, we use the fact that a translation in Fourier space may be expressed as a phase ramp in real space.
Thus, following the detailed presentation of \cite{qian_structured_2023}, transforming each component into real space allows for a minimization to yield the sub-pixel phase vector using the explicit cross-correlation over $\mathbf{r} = (x, y)^\mathrm{T}$
\begin{equation}
\begin{split}
    CC_\mathrm{(n,m)}(\mathbf{q}^{\prime}_\mathrm{sub}) =& \frac{1}{\sum_\mathrm{\mathbf{r}}|S_\mathrm{(0,0)}(\mathbf{r})|^2}\\
    &\cdot \sum_\mathrm{\mathbf{r}}A_\mathrm{(n,m)}S^*_\mathrm{(0,0)}(\mathbf{r})S_\mathrm{(n,m)}(\mathbf{r})\\
    &\cdot e^{i(\varphi_{0, \mathrm{(n,m)}}+2\pi\mathbf{r}(\mathbf{q}_\mathrm{sub}-\mathbf{q}^\prime_\mathrm{sub})))}\\
    =& \frac{\sum_\mathrm{\mathbf{r}}S^*_\mathrm{(0,0)}(\mathbf{r})D_\mathrm{(n,m)}(\mathbf{r})e^{i2\pi\mathbf{r}(\mathbf{q}_\mathrm{sub}-\mathbf{q}^\prime_\mathrm{sub})}}{\sum_\mathrm{\mathbf{r}}|S_\mathrm{(0,0)}(\mathbf{r})|^2}.
\end{split}
\end{equation}
Here, $S^*_\mathrm{(0,0)}$ is the complex conjugated zeroth-order real-space sample replica, $S_\mathrm{(n,m)}$ is the real-space sample replica of order $(n,m)$, shifted by the integer component of $\mathbf{q}$.
The vector $\mathbf{q^\prime_\mathrm{sub}}$ represents a sub-pixel shift guess used in minimization.
The value of $\mathbf{q^\prime_\mathrm{sub}}$ for which $|CC_\mathrm{(n, m)}|$ is maximal yields the desired sub-pixel shift $\mathbf{q_\mathrm{sub}}$.\newline
Having estimated the sub-pixel shift vector, the parameters $A_\mathrm{(n,m)}$ and $\varphi_\mathrm{0, (n,m)}$ can be obtained by complex linear regression:
\begin{align}
\begin{split}
    A_\mathrm{(n,m)} &= \left| \frac{\sum_\mathrm{\mathbf{r}} A_\mathrm{(n,m)}S^*_\mathrm{(0,0)}(\mathbf{r})S_\mathrm{(n,m)}(\mathbf{r})e^{i\varphi_0}}{\sum_\mathrm{\mathbf{r}}|S_\mathrm{(0,0)}(\mathbf{r})|^2} \right|\\
    &= \left| \frac{\sum_\mathrm{\mathbf{r}} S^*_\mathrm{(0,0)}(\mathbf{r})D_\mathrm{(n,m)}(\mathbf{r})}{\sum_\mathrm{\mathbf{r}}|S_\mathrm{(0,0)}(\mathbf{r})|^2} \right|
\end{split}
\end{align}
\begin{align}
\begin{split}
    \varphi_0 &= \mathrm{arg}\left[ \frac{\sum_\mathrm{\mathbf{r}} A_\mathrm{(n,m)}S^*_\mathrm{(0,0)}(\mathbf{r})S_\mathrm{(n,m)}(\mathbf{r})e^{i\varphi_0}}{\sum_\mathrm{\mathbf{r}}|S_\mathrm{(0,0)}(\mathbf{r})|^2} \right]\\ &= \mathrm{arg}\left[ \frac{\sum_\mathrm{\mathbf{r}} S^*_\mathrm{(0,0)}(\mathbf{r})D_\mathrm{(n,m)}(\mathbf{r})}{\sum_\mathrm{\mathbf{r}}|S_\mathrm{(0,0)}(\mathbf{r})|^2} \right].
\end{split}
\end{align}
Following this estimation, \cite{gustafsson_three-dimensional_2008} redefines the OTF such that these parameters are absorbed into it.
Accordingly, defining
\begin{equation}
    O^\prime_\mathrm{(n,m)}(u,v) = A_\mathrm{(n,m)}e^{i\varphi_{0,\mathrm{(n,m)}}}O(u,v)
\end{equation}
enables the population of the super-resolved Fourier space through a generalized Wiener filter
\begin{equation}\label{eq:wienerfilter}
\begin{split}
    \Tilde{S}(u,v)  =& \frac{1}{\sum_\mathrm{n,m}|O^\prime_\mathrm{(n,m)}(u+nq, v+mq)|^2 + \epsilon^2}\\
    &\cdot \sum_\mathrm{n,m}O^{\prime *}_\mathrm{(n,m)}(u+nq, v+mq)\Tilde{S}_\mathrm{(n,m)}(u,v)\\
    &\cdot O^\prime_\mathrm{(n,m)}(u+nq, v+mq)\\
    =& \frac{\sum_\mathrm{n,m}O^{\prime *}_\mathrm{(n,m)}(u+nq, v+mq)\Tilde{C}_\mathrm{(n,m)}(u,v)}{\sum_\mathrm{n,m}|O^\prime_\mathrm{(n,m)}(u+nq, v+mq)|^2 + \epsilon^2}
\end{split}
\end{equation}
where $\Tilde{S}_\mathrm{(n,m)}(u,v)=\mathscr{F}\{S(x,y)\}_\mathrm{(n,m)}(u,v)$ are the separated sample replicas in Fourier space, shifted to their true position.
The modulated OTFs are shifted according to the shift vector to retain their effects on their associated components.
The parameter $\epsilon$ is a positive, empirically chosen regularization parameter.
To control high-frequency noise and artifacts, a cosine apodization with a width determined by the frequency interval in which information was generated is applied to Eq.\ \eqref{eq:wienerfilter}.
The inverse Fourier transformation of Eq.\ \eqref{eq:wienerfilter} yields the final real-space super-resolution reconstruction.

\section{Methodology}
\subsection{Experimental Setup for Test Pattern Acquisition}\label{sec:methodsetup}
\noindent As discussed in Sec.\ \ref{sec:additionalinfo}, the reconstruction method requires a high-visibility illumination structure imposed on the sample projection.
As had been established since \cite{suleski_generation_1997}, a binary phase grating with a duty cycle of $1/3$ and an induced phase shift of $2\pi/3$ produces a self-image at the fractional Talbot distance of $Z_\mathrm{T}/6$, where $Z_\mathrm{T}$ denotes the Talbot distance.
This has been experimentally demonstrated by \cite{gustschin_high-resolution_2021} for highly coherent X-ray radiation and a 2D binary phase grating.
For our experiment, a two-dimensional silicon phase grating with a period of $\qty{5}{\micro\meter}$ was used as a wavefront marker at the P05 beamline \cite{stock_p05_2014,wilde_micro-ct_2016} at PETRA III at the Deutsches Elektronen-Synchrotron (DESY, Hamburg, Germany) at an X-ray energy of $\qty{18}{\kilo\electronvolt}$. 
The grating is designed to induce a phase shift of $2\pi/3$ at \qty{20}{\kilo\electronvolt}. 
A simulation found a fractional Talbot distance of \qty{121}{\milli\meter}. 
Since the method also requires a translation of the structured illumination on the scale of the pattern period, the grating was mounted on a piezo-electric actuator.
To determine the visibility of the Talbot illumination, the intensity variation recorded by each pixel during the stepping sequence is analyzed to obtain a stepping profile for each pixel.
The visibility change is then computed from this profile using Eq.\ \ref{eq:visibility}.
The resulting visibility map is shown in Supplementary Figure 7 (d).
The visibility value for the region discussed in Sec.\ \ref{sec:srdemo} corresponds to the mean over this visibility map. \newline
The scintillation screen is located at the position of highest theoretical visibility, when including effects from the convolutional kernel $K(x,y)$ in Eq.\ \eqref{eq:realspaceintensity}.
The scintillator is coupled to an sCMOS pixelated detector \cite{stock_characterization_2014} using a $5\times$ magnifying objective, resulting in an effective pixel size of \qtyproduct{1.28x1.28}{\micro\meter}.
To determine the PSF of the system, a circular aperture with a diameter of \qty{1}{\mm} was positioned as close to the detector as possible, and a series of $100$ images was acquired.
This was followed by the same number of dark-current and flat-field acquisitions.
The exposure time was \qty{60}{\milli\second} for all measurements.
The images were averaged to increase statistical reliability, and the PSF was reconstructed as detailed in \cite{forster_single-shot_2025}.\newline
The experiment was conducted using a resolution test pattern (model X500-200-30, Xradia, Pleasanton, USA; now ZEISS, Oberkochen, Germany) was positioned $\qty{2}{\milli\meter}$ upstream of the scintillator screen.
For the depiction in Figure \ref{fig:first_plot} and Figure \ref{fig:second_plot}, a set of 25 images was acquired in a $5\times5$ raster, with step sizes of \qty{1}{\um} in both dimensions.
This sampling strategy enables separation of the first- and second-harmonic orders under conditions of maximal variation in grating position across images.
At each position, four exposures of $\qty{90}{\milli\second}$ were averaged.
This procedure was performed once with the sample in the beam path and once without to obtain a flat-field reference.

\subsection{Experimental Setup for Tomographic Acquisition}\label{sec:tomosetup}
\noindent The tomographic acquisition used for super-resolution reconstruction has previously been reported on by \cite{john_near-perfect_2026}.
It was performed at the GINIX setup \cite{kalbfleisch_gottingen_2011,salditt_x-ray_2026} of the beamline P10 at DESY using an Eiger X 4M photon-counting detector by DECTRIS AG (Baden-Daettwil, Switzerland) with a physical pixel size of $\qtyproduct{75x75}{\um}$.
As this beamline relies on a small source spot emerging from a waveguide \cite{kalbfleisch_gottingen_2011}, large geometric magnification is central to the imaging setup.
Accordingly, the detector was positioned at a distance of $\qty{5090}{\milli\meter}$ from the waveguide.
With the sample positioned at a distance of $\qty{260}{\milli\meter}$ from the waveguide, a magnification of $\qty{19.6}{}$ was achieved, yielding an effective pixel size of \qtyproduct{3.84x3.84}{\um} in the sample plane. \newline
A two-dimensional binary phase grating, as described by \cite{gustschin_high-resolution_2021}, with a design period of $\qty{10}{\micro\meter}$ was positioned at a distance of $\qty{100}{\milli\meter}$ from the waveguide's exit.
It is designed to affect a $2\pi/3$ phase shift at an energy of $\qty{10}{\kilo\electronvolt}$.
The X-ray energy used in the experiment was $\qty{8}{\kilo\electronvolt}$.
Through a magnification of $\qty{50.9}{}$, its effective grating period in the detector plane was $\qty{509}{\micro\meter}$.
In this setup, each grating period was sampled by $6.8$ detector pixels, corresponding to a grating-to-detector frequency ratio of $3.4$.
To demonstrate that the methodology can be extended to tomography with a similar resolution increase and for higher harmonic orders, the relationship between the Nyquist and grating frequencies needs to be adjusted.
Hence, the data set is downsampled, increasing the detector pixel size to \qtyproduct{114x114}{\um} and adjusting the ratio between the sampling and grating frequencies to $2.2$.
Consequently, both the first and second harmonic orders are firmly within the native Nyquist frequency.
This does not restrict the generality of the demonstration, 
as a different magnification of the grating can achieve the appropriate frequency in the detector plane. \newline
The tomography was acquired over a $\qty{360}{\degree}$ angular range, with $250$ equidistant angular projections.
The exposure time was set to $\qty{1}{\second}$ with an energy threshold of $\qty{4.4}{\kilo\electronvolt}$.
A total of $26$ steps was acquired in the scheme detailed in \cite{gustschin_high-resolution_2021} with $a=1$ and $b=5$, of which only $25$ were used to reconstruct up to the second harmonic order.
At each grating step, a full tomography was acquired.

\subsection{Reconstruction}\label{sec:reconstruction}
\noindent The super-resolution reconstruction for each projection is performed using a custom Python framework designed based on the descriptions in Sec.\ \ref{sec:additionalinfo} and \ref{sec:recoveringadditionalinformation}.
The tomographic reconstruction is computed with the software X-AID (MITOS, Munich, Germany). \newline
For both the test pattern acquisition and the tomography, the reconstruction procedure was performed independently for the sample and reference datasets.
The precise grating positions used to populate the matrix in Eq.\ \eqref{eq:separation_matrix_full} were estimated using the non-iterative approach described by \cite{wicker_non-iterative_2013}.
The final image for each projection was obtained by dividing the reconstructed sample image by the corresponding reference, yielding a properly scaled result.
After the test pattern reconstruction, the extended pass-band is sampled with an effective pixel size of $\qtyproduct{0.64x0.64}{\um}$.
For analysis, the image was slightly rotated using a third-order spline interpolation to align the horizontal and vertical line patterns with their respective axes.
Further, aliasing effects had to be compensated in the test pattern acquisition.
If the PSF of an imaging system does not sufficiently suppress information outside the pass-band, aliasing artifacts may significantly influence the acquired image.
In the case described in Sec.\ \ref{sec:demoadditional}, the effective detector pixel pitch is $\qty{1.28}{\micro\meter}$.
Given the grating design period of $\qty{5}{\micro\meter}$, the second-order harmonic is only marginally unresolved, resulting in an aliasing artifact in the Fourier domain of the acquisition.
In the case presented here, this can be corrected by periodically shifting the Fourier image.
This approach effectively suppresses dominant aliasing artifacts, provided two conditions are met.
First, the frequency may only be unresolved by a marginal distance in frequency space, thereby minimizing the required pixel shift.
Second, the peak must be well modulated, ensuring that its information amplitude is sufficiently dominant over the unaliased sample frequencies. \newline
For the tomographic dataset, optimal synthetic references for each image were computed from a set of sample-free acquisitions, as described by \cite{van_nieuwenhove_dynamic_2015}.
In the super-resolution reconstruction and parameter estimation procedure, the PSF is assumed to be box-shaped.
Defective pixels were corrected using a median filter.
After reconstruction, the extended pass-band is sampled by an effective pixel size of $\qty{2.04}{\micro\meter}$. 
The three-dimensional volume was reconstructed using filtered back-projection with a Ram-Lak filter in the software X-AID.

\subsection{Resolution Assessment}\label{sec:resolutionclassification}
\noindent To determine the achievable resolution gain, a test pattern was imaged as detailed in Sec.\ \ref{sec:methodsetup}.
The resolution achieved for the test pattern (see Figure \ref{fig:second_plot}) is quantified in terms of line-pair modulation defined by the visibility
\begin{equation}\label{eq:visibility}
    M = \frac{I_\mathrm{max} - I_\mathrm{min}}{I_\mathrm{max} + I_\mathrm{min}},
\end{equation}
where $I$ denotes the detected intensity.
The maximal and minimal intensity values $I_\mathrm{max}$ and $I_\mathrm{min}$ were taken from the associated line-pair region after averaging along the axis perpendicular to the line-pattern.
A line-pair is considered resolved when the modulation exceeds the $\qty{10}{\percent}$ threshold \cite{noauthor_digital_2024}. \newline
The resolution gain achieved in the tomographic reconstruction was determined using Fourier ring correlation \cite{saxton_correlation_1982}.
For our purpose, a feature is considered resolved if its correlation exceeds the half-bit criterion \cite{van_heel_fourier_2005}.

\section{Results}
\subsection{Demonstrating the Presence of Additional Information}\label{sec:demoadditional}
\begin{figure*}[!t]
\centering
\includegraphics[width=440pt]{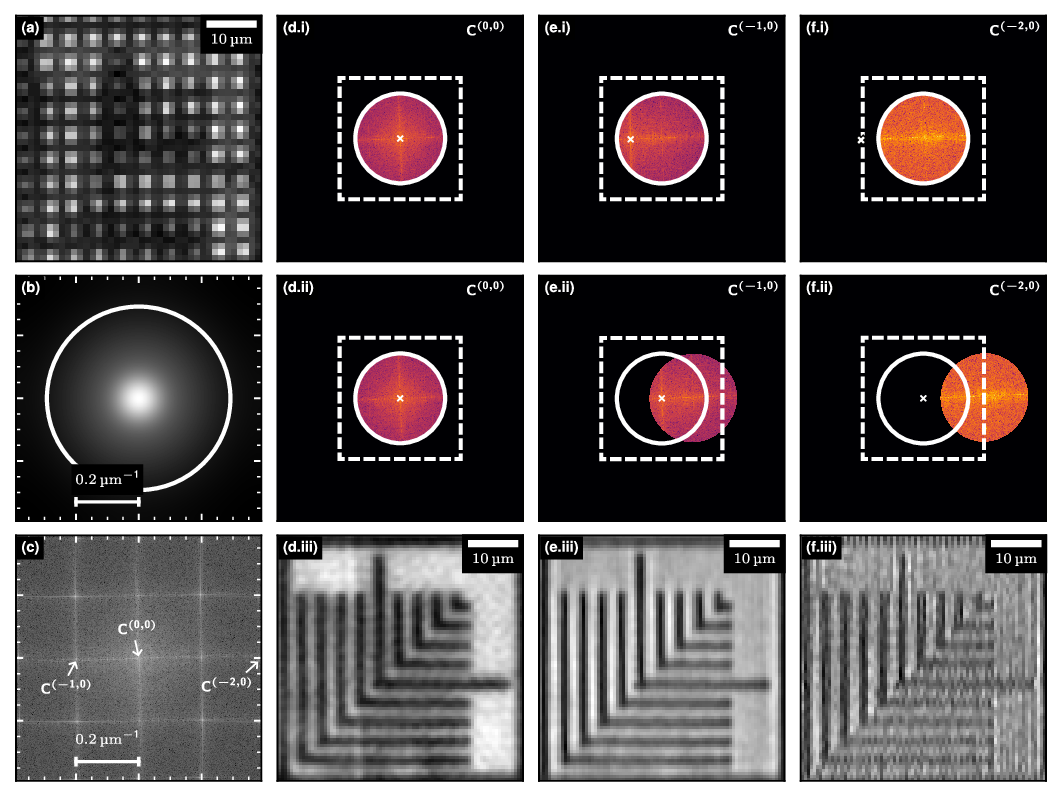}
\caption{Descriptive overview of the frequency landscape and its implications.
    (a) Exemplary region of the periodically modulated acquisition.
    (b) Modulation transfer function.
    The white circle indicates the region of support of $>\qty{10}{\percent}$ modulation.
    (c) The logarithmic absolute value of the discrete Fourier transform of a single acquired image under structured illumination.
    Selected peaks in the Dirac comb are labeled.
    The second-order harmonic is marked at its aliased position.
    Each column (d--f) discusses an associated separated component marked in (c).
    The components are separated and zero-padded to twice their size to satisfy the Nyquist criterion for the super-resolved pass-band.
    The rows (i-iii) depict the same state for each component.
    (i) Separated components.
    Their information contribution is assumed to be zero outside of the region of support inferred from (b) and marked by the solid white circle.
    The additional constraint of the Nyquist frequency is shown by the white dashed square.
    The origin of each sample replica is marked by a white cross.
    (ii) The components of (i) shifted to their true position.
    The marked origin of the replica coincides with the origin of Fourier space.
    The shifted information exceeds both the previous region of support and the Nyquist limit.
    (iii) The absolute value of the inverse Fourier transform of (ii).
    Evidently, the feature modulation increases along the dimension of the shift vector.}
\label{fig:first_plot}
\end{figure*}

\noindent As a first step, we demonstrate the presence of additional information in an image of a resolution test pattern modulated by a structured illumination.
Figure \ref{fig:first_plot} (a) displays an exemplary region in the modulated sample acquisition described in Sec.\ \ref{sec:methodsetup}.
The mean detected visibility over the full test pattern region is $\qty{23.9}{\percent}$ following Eq.\ \eqref{eq:visibility} and the procedure detailed in Sec.\ \ref{sec:methodsetup}.
(b) shows the modulation transfer function (MTF), with the $\qty{10}{\percent}$ threshold marked by a white circle.
(c) illustrates the logarithmic absolute value of the Fourier transform of the test pattern region.
The zeroth-order component $C^{(0,0)}$ is centered around the origin of Fourier space.
Thus, this component contributes information equivalent to an unmodulated native acquisition.
The component $C^{(-1,0)}$ is centered around the peak of the Dirac comb positioned at the first-order harmonic of the grating in the respective dimension.
The $C^{(-2,0)}$ component exhibits a more complex association.
In the undetected intensity field, the center of the sample replica is at the corresponding integer multiple of the grating frequency.
However, the detector samples this continuous signal with a finite number of discrete pixels.
As this sampling is insufficient to resolve the second-order harmonic peaks, aliasing occurs, i.e., the peaks are wrapped to the opposite side of the pass-band.\newline
Figure \ref{fig:first_plot} is in part organized as a grid: columns (d--f) correspond to components $C^{(0,0)}$, $C^{(-1,0)}$, and $C^{(-2,0)}$, respectively.
Rows (i--iii) show (i) separated components, (ii) their true frequency positions, and (iii) a region-of-interest of their inverse Fourier transforms.
The native resolution limits are marked in (i) and (ii) to illustrate how these components provide additional information.
In visible light microscopy, the primary limitation to resolution is the diffraction limit $\frac{0.61\lambda}{\mathrm{NA}}$ \cite{rayleigh_xxv_1881,born_principles_2017}, where $\lambda$ is the imaging wavelength and $\mathrm{NA}$ is the numerical aperture.
While this constraint also holds in X-ray imaging, other effects impose more stringent limits on resolution at wavelengths of the order of $\qty{1}{\angstrom}$.
One such effect is the OTF of the imaging system itself.
It down-modulates high spatial frequencies, thereby curbing resolution.
This region of support, defined as the region where the OTF permits intensities above the noise threshold, is indicated by the solid white outline in rows (i--ii).
Further, the maximum sampling frequency imposed by the detector pixel size restricts the resolution to the Nyquist limit, which is indicated by a dashed square in (i--ii).
Overcoming this limitation is of great interest since imaging setups are typically well-optimized.
Accordingly, a super-resolution method that merely compensates the OTF within the existing region of support offers limited benefit.
Subfigure (d.i) displays the separated zeroth-order component, bounded by both the OTF and the Nyquist frequency.
The absolute value of its inverse Fourier transform returns the unmodulated native detector image depicted in (d.iii).
The aliased Fourier-space component in (f.i) is mapped to its unaliased position by an integer-pixel translation in the Fourier coordinate space, before being shifted to its true position shown in (f.ii).
This is justified in Sec.\ \ref{sec:reconstruction}. \newline
Notably, high-frequency information, which lies outside the imposed resolution limit for the native image, is firmly within the region of support of higher-order components, allowing the population of an extended frequency space.
The real-space implications are illustrated in row (iii), which presents the absolute value of the inverse Fourier-transformed components in (ii).
The observed increase in pattern modulation along the direction of the component shift vector confirms the presence of additional high-frequency information.

\subsection{Demonstration of Super-Resolution}\label{sec:srdemo}
\begin{figure*}[!t]
\centering
\includegraphics[width=440pt]{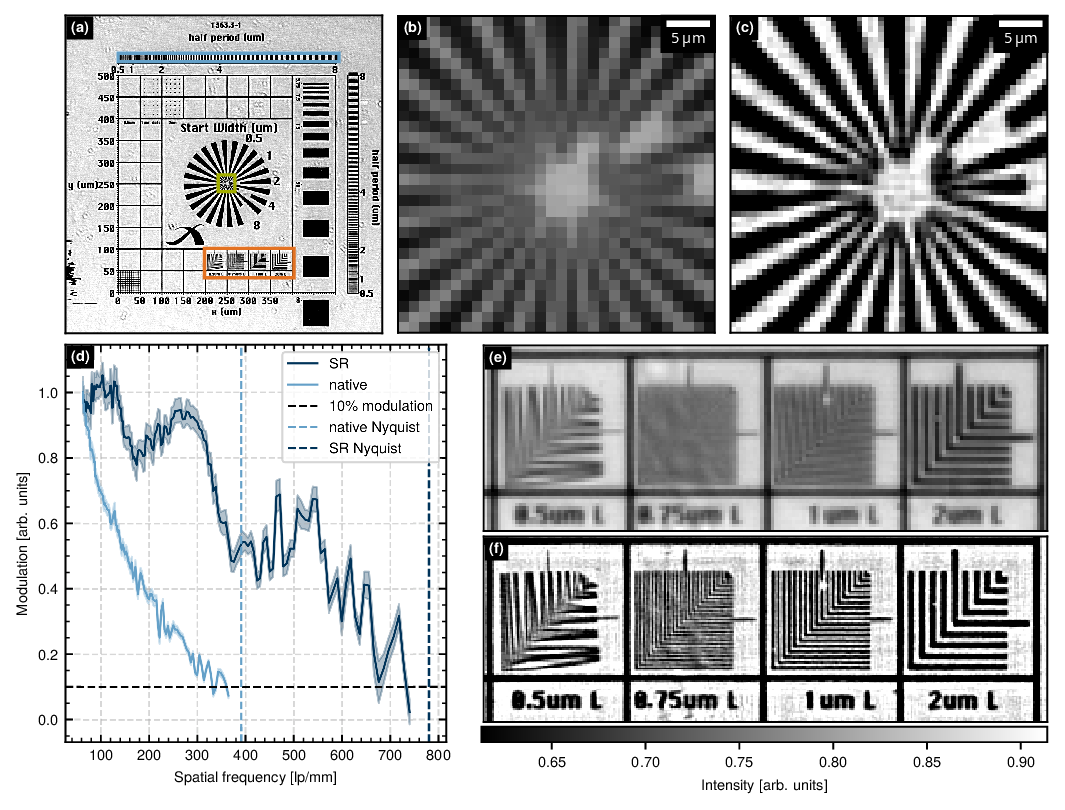}
\caption{Reconstruction and analysis for a resolution test pattern projection.
    (a) Reconstructed super-resolved projection of the pattern.
    The height map is normalized by a reference reconstruction.
    (b) Native image of the center region of the Siemens star in the middle of the test pattern marked in green in (a).
    (c) Second-order super-resolved image of the same region.
    (d) Modulation of the line pattern marked in light blue in (a) as a function of spatial frequency.
    The error margins indicate the propagated standard deviation of the noise.
    A modulation of $\qty{10}{\percent}$ is chosen as the resolution criterion.
    The modulation in the reconstructed super-resolution (SR) images is superior across the entire frequency range and extends beyond the native Nyquist limit.
    (e) Native image of the L-shaped resolution target, marked in orange in (a).
    The \qty{0.5}{\micro\meter}, \qty{0.75}{\micro\meter} and \qty{1}{\micro\meter} features are not resolved.
    The \qty{2}{\micro\meter} features are well resolved.
    (f) Super-resolved image of the L-shaped resolution target marked in orange in (a).
    The \qty{2}{\micro\meter}, \qty{1}{\micro\meter} and \qty{0.75}{\micro\meter} features are resolved.
    The \qty{0.5}{\micro\meter} features remain unclear.}
\label{fig:second_plot}
\end{figure*}

\noindent The absolute value of the inverse Fourier transform of the separated zeroth order yields the unmodulated native detector image, which is used as a baseline to quantify the achieved resolution enhancement.
In visible-light structured-illumination microscopy, the mean image over all pattern steps is typically used as the widefield equivalent \cite{wen_spectrum-optimized_2023}.
However, to remove low-frequency modulation, this approach assumes that the pattern's translation is perfectly equidistant.
It is thus more appropriate to use the $C^{(0,0)}$ components, thereby alleviating the imposition of equidistance under the assumption that the separation matrix is well estimated. 
The precise steps taken in the reconstruction are detailed in Sec.\ \ref{sec:recoveringadditionalinformation}.\newline
As the reconstruction requires parameter estimation of an intensity field independent of detection, the effects of detection, specifically the point spread function (PSF), must be known precisely.
It is reconstructed from an image of a circular aperture, as described by \cite{forster_single-shot_2025}, and adequately modeled by the sum of two Gaussians with the mean widths of $\sigma_1=\qty{0.96}{\micro\meter}$ and $\sigma_2=\qty{3.32}{\micro\meter}$.\newline
The super-resolution test pattern projection, reconstructed up to the second-order harmonic, is shown in Figure \ref{fig:second_plot} (a).
The full test pattern projection for both the native and super-resolved cases is depicted in Supplementary Figures 4 and 5, respectively.
A comparison between first- and second-order reconstruction is depicted in Supplementary Figure 6.
The height map in Figure \ref{fig:second_plot} is normalized by a reference reconstruction.
Subfigures (a--c) and (e--f) are all plotted on a common intensity scale.
The line pattern marked by the light blue box is used for the modulation assessment in (d), the green square indicates the region-of-interest displayed in (b) and (c).
Orange indicates the region used for the comparison in (e) and (f).
An initial qualitative assessment can be made for the two enlarged regions of interest displayed in (b--c) and (e--f).
When comparing the native scan in (b) to the super-resolved image in (c), the center of the Siemens star shows a stark difference in modulation.
Image (c) further demonstrates that the extended pass-band is populated isotropically, as the resolution gain shows no apparent directionality.
The \qty{2}{\micro\meter} line pattern in (e--f) is clearly resolved in both the native image and the reconstruction, with a native modulation at this resolution of $\qty{28.9}{\percent}$ and a reconstructed modulation of $\qty{92.4}{\percent}$.
The resolution gain is more readily apparent for the \qty{1}{\micro\meter} pattern, as its feature size lies below the native effective pixel size of \qty{1.28}{\micro\meter}.
Accordingly, the native image does not resolve the pattern, while the reconstruction shows a modulation of $\qty{52.3}{\percent}$.
This confirms the possibility of reconstructing information past the Nyquist limit of the imaging system.
Similarly, the \qty{0.75}{\micro\meter} line pattern is not resolved in the native scan, yet, although less so than the former, it is modulated in the reconstruction, which at this linewidth demonstrates a modulation of $\qty{20.3}{\percent}$.
Contrary, the $\qty{0.5}{\micro\meter}$ pattern is not resolved and further exhibits pronounced distortions.
While this pattern irregularity may be exacerbated by Moir\'e artifacts, both the native and the reconstructed image show the same deformation morphology, so it is likely due to physical damage of the pattern.
The quantitative modulation values in (d) are determined from the line pattern marked by the blue box in (a).
For our purpose, a structure of a certain size is considered resolved if the modulation of the line pattern at this size exceeds \qty{10}{\percent}\ \cite{noauthor_digital_2024}.
Based on this criterion, the native resolution is \qty{331}{lp/\milli\meter}, while the reconstructed resolution is \qty{733}{lp/\milli\meter}.
Hence, the resolution is increased by a factor of $2.2$.
The error margin is computed as the propagated error originating from the standard deviation of noise along the line pattern width.
The effective pixel size in the reconstructed image is decreased to $\qty{0.64}{\micro\meter}$ to satisfy the Nyquist criterion for the newly generated information components.
As the line pattern frequency approaches the extended sampling frequency, the modulation values fluctuate depending on whether the modulation peak is distributed between two pixels or contained within a single pixel.

\subsection{Super-Resolution Tomography}\label{sec:srtomo}
\begin{figure*}[!t]
\centering
\includegraphics[width=440pt]{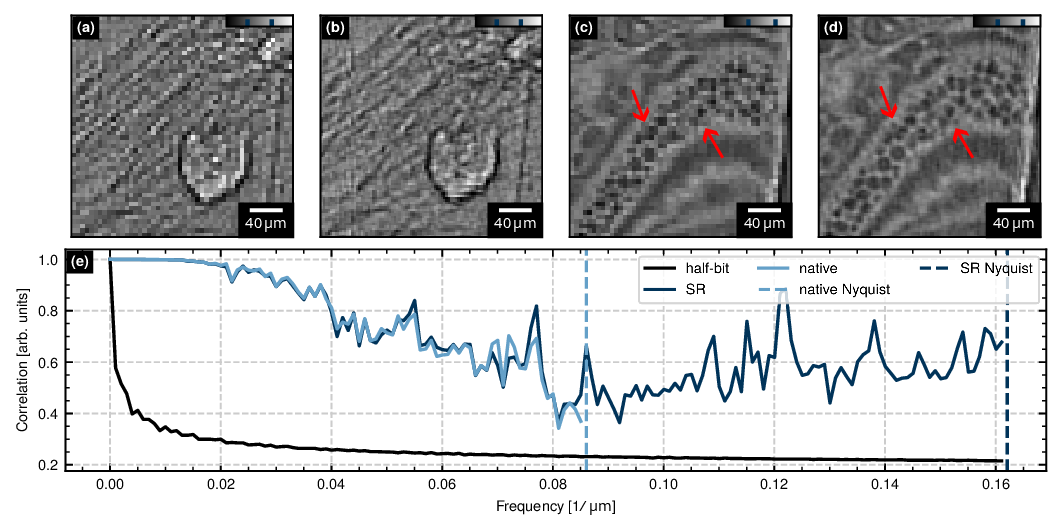}
\caption{Native and super-resolved tomographic reconstruction.
    (a) Native downsampled projection.
    (b) Second-order super-resolution reconstruction of the downsampled projection.
    The colorbars in (a) and (b) are marked at $0.7$ and $1.1$ in arbitrary units of normalized intensity.
    (c) Region-of-interest in a tomographic slice of the native downsampled reconstruction.
    (d) Region-of-interest in a tomographic slice of the second-order super-resolution reconstruction of the downsampled projections.
    The red arrows in the tomographic reconstructions highlight features that are indistinguishable in the native image in (c) but become discernible in the super-resolved image in (d). 
    The two ticks in the colorbars in (c) and (d) mark the reconstructed linear attenuation at $0$ and $10$ in units of \qty{}{\per\milli\meter}.
    (e) Fourier ring correlation computed on (a) and (b) and respective neighboring projections.
    Features are considered resolved as long as their correlation exceeds the half-bit criterion \cite{van_heel_fourier_2005}.
    In the super-resolution (SR) reconstruction, the correlation is sustained beyond the Nyquist limit of the native image. }
\label{fig:tomo_plot}
\end{figure*}

\noindent Beyond the resolution gain quantified on a projection image in Sec.\ \ref{sec:srdemo}, we demonstrate the applicability of our methodology for X-ray tomography.
To this end, a tomographic dataset of a mouse skin was provided by \cite{john_near-perfect_2026}.
The data acquisition and reconstruction are detailed in Sec.\ \ref{sec:tomosetup}.
Figure \ref{fig:tomo_plot} (a) shows an exemplary region of a native downsampled projection.
Figure \ref{fig:tomo_plot} (b) shows the super-resolution reconstruction.
Example regions from the tomographic reconstruction of both the native and super-resolved scans are depicted in (c) and (d), respectively.
Under qualitative assessment of both the projection and the tomogram, the super-resolved reconstruction enables differentiation of features that are indistinguishable in the native scan.
This is exemplified by the features highlighted by the red arrows in Figure \ref{fig:tomo_plot} (c) and (d).
In the native tomographic reconstruction in (c), these features are indiscernible, while they are distinguishable in the super-resolved reconstruction in (d).
This is quantified by the Fourier ring correlation \cite{saxton_correlation_1982} shown in subfigure (e).
The correlation is computed between two neighboring projections, assuming that the angular difference is small enough to allow for correlation of the sample information.
Up to the Nyquist frequency of \qty{0.086}{\per\um} calculated from the native pixel size, the correlation of the super-resolved reconstruction follows the native correlation.
The fundamental frequency of the structured illumination is \qty{0.038}{\per\um}, extending the recoverable information to \qty{0.124}{\per\um}, as confirmed by the correlation function.
The second-order harmonic has a frequency of \qty{0.077}{\per\um}, which results in additional information up to \qty{0.163}{\per\um}.
This is demonstrated again by the sustained correlation function over this interval.
It supports the qualitative improvements seen in Figure \ref{fig:tomo_plot} (a--d), confirming that our super-resolution approach is effectively applicable for tomographic imaging.

\section{Discussion \& Conclusion}
\noindent This resolution increase beyond the Nyquist limit is particularly relevant for photon-counting detectors, for which pixel sizes are presently limited to about $\qty{50}{\micro\meter}$, since charge sharing becomes increasingly significant as pixel dimensions decrease \cite{chmeissani_performance_2001}.
Our approach enables surpassing this limit, as demonstrated in Sec.\ \ref{sec:srtomo}.
In contrast to super-sampling methods such as proposed by \cite{ehn_x-ray_2016}, our method does not rely on the performance of a deconvolution, but rather expands the pass-band independently of detection.
This pass-band could be expanded further, incorporating unresolved higher orders into the reconstruction, see \cite{gustafsson_nonlinear_2005}.
Reconstructing the extended pass-band used in this work based on information carried by the first and the second harmonic of a square grating requires 25 acquisition steps.
Compared to the full-field structured-illumination approach formulated by \cite{gunther_full-field_2019}, our method achieves a comparable resolution gain while reducing the necessary acquisition steps by a factor of $4$.
As the maximal theoretically reconstructable frequency depends on the grating frequency, the same resolution gain can be achieved by reconstructing only the first order, provided the grating frequency is doubled.
Thus, our proposed method would yield similar results with only nine translations.
This approach may even surpass the resolution gain demonstrated for second-order harmonic reconstruction, as the first-order peak amplitude is higher (see Eq.\ \eqref{eq:illspectrum}).
Adopting high-frequency hexagonal grating structures would allow for an isometric frequency-space population with only seven steps for the first-order reconstruction.
This reduces the acquisition time by more than a factor of $14$ compared to the approach by \cite{gunther_full-field_2019}.\\
Like in their work, the acquisition time does not scale with the size of the field of view (FOV).
This differentiates our approach from classical STXM, which are typically limited to sample sizes on the order of \qty{5}{\micro\meter} \cite{aidukas_high-performance_2024} to a few \qty{100}{\micro\meter} \cite{kilcoyne_interferometer-controlled_2003}.
This restriction arises from factors including the acquisition time, the limited depth of focus \cite{aidukas_high-performance_2024}, and the travel range of the piezoelectric stages used for scanning \cite{kilcoyne_interferometer-controlled_2003}.
In contrast, our method is well-suited for high-resolution, large FOV imaging in applications where resolution gain from magnifying optics is unsuitable.
The required gratings can be fabricated with more than \SI{10}{\cm} in diameter, even at tens of kiloelectronvolt X-ray energy.
Therefore, the FOV is X-ray beam size-limited at synchrotrons.
However, the beam size may be matched to the grating size, e.g., using a transfocator \cite{vaughan_x-ray_2011} or a Bragg-magnifier \cite{boettinger_x-ray_1979}.
Consequently, large FOV imaging, such as demonstrated on rodent cochleas by \cite{schaeper_3d_2025}, becomes feasible at higher resolutions due to our method.
The reliance on as well as the complications from stitching multiple tomograms, which typically would be required when moving to a setup with larger geometric magnification, is thereby also alleviated.\newline
Nevertheless, the number of required acquisition angles in tomography increases as the super-resolved projections have more pixels.
Although this is true independently of the imaging method used, the necessity for multiple acquisitions per projection results in longer acquisition times and substantially larger datasets.
Moreover, inaccuracies in the stepping parameters due to positioning errors heavily influence the reconstruction performance as depicted in Supplementary Figure 8.
Consequently, iterative phase-optimization techniques \cite{wicker_phase_2013} are indispensable for determining the actual stepping and phase shift in such situations to improve reconstruction effectiveness and reduce artifacts, as discussed in Supplementary Figure 7.
This is especially important within a tomographic framework.
Consequently, reconstructing these datasets requires intensive computational effort.
However, this can be addressed by developing a GPU-accelerated reconstruction framework, which is justified as all processing steps are well parallelizable.
This work is currently ongoing.\\
Another very interesting aspect of our method is that data acquisition is fully compatible with phase-contrast and dark-field imaging using unified modulated pattern analysis \cite{zdora_x-ray_2017} (UMPA) or Fourier space methods \cite{bennett_gratingbased_2010,wen_spatial_2008}.
One key advantage of our proposed method is that component separation in our formulation can enable higher resolution in contrast to traditional Fourier space approaches, as discussed by \cite{he_application_2019}.
If the component separation is well optimized through phase-optimization algorithms \cite{shroff_phase-shift_2009,wicker_non-iterative_2013,wicker_phase_2013,cao_inverse_2018}, the resolution increase determined in \cite{he_application_2019} may even be exceeded. 
Further, for propagation-based phase contrast \cite{paganin_simultaneous_2002}, our method may alleviate the resolution loss introduced by the low-pass filter.\\
In conclusion, rapid multimodal super-resolution X-ray microscopy and micro-tomography of large fields-of-view become feasible with our method, which employs structured X-ray illumination and Fourier decomposition.
Since only a structured intensity distribution in the detector plane is required to retrieve super-resolved images, the restriction on partially spatially coherent X-rays can be alleviated by placing an absorption grating close to the detector.
This extends the applicability of our approach to a wide range of laboratory-based X-ray sources.
Hence, we believe that our method will be widely adopted for non-destructive testing and biomedical imaging, as it alleviates pixel-size limitations in detectors and sample-size restrictions.

\section*{Data and code availability}
\noindent The raw data and code used for the super-resolution reconstruction of the test pattern may be obtained from the authors upon reasonable request. The raw data used for the tomographic reconstruction were provided by John et al.\ \cite{john_near-perfect_2026} and are available upon reasonable request to John et al.

\section*{Acknowledgments}
\noindent We gratefully acknowledge the Helmholtz-Zentrum Hereon and the Deutsches Elektronen-Synchrotron for providing the required research infrastructure under the proposals BAG-20240009 and BAG-20240010.
We also extend our gratitude to the beamline staff, namely Jörg Hammel and Felix Beckmann, for their continuous assistance.
We further thank Prof.\ Julia Herzen for her assistance in organizing the beamtimes at DESY and Dominik John for fruitful discussions and for providing the tomography dataset.
Benedikt Günther and Lennart Forster acknowledge funding by the Deutsche Forschungsgemeinschaft (513827659).
All authors acknowledge funding from the Centre for Advanced Laser Applications.

\bibliographystyle{IEEEtran}
\bibliography{bibliography}

\vfill

\pagebreak

\setcounter{figure}{3}

\begin{figure*}[!t]
\centering
\includegraphics[width=440pt]{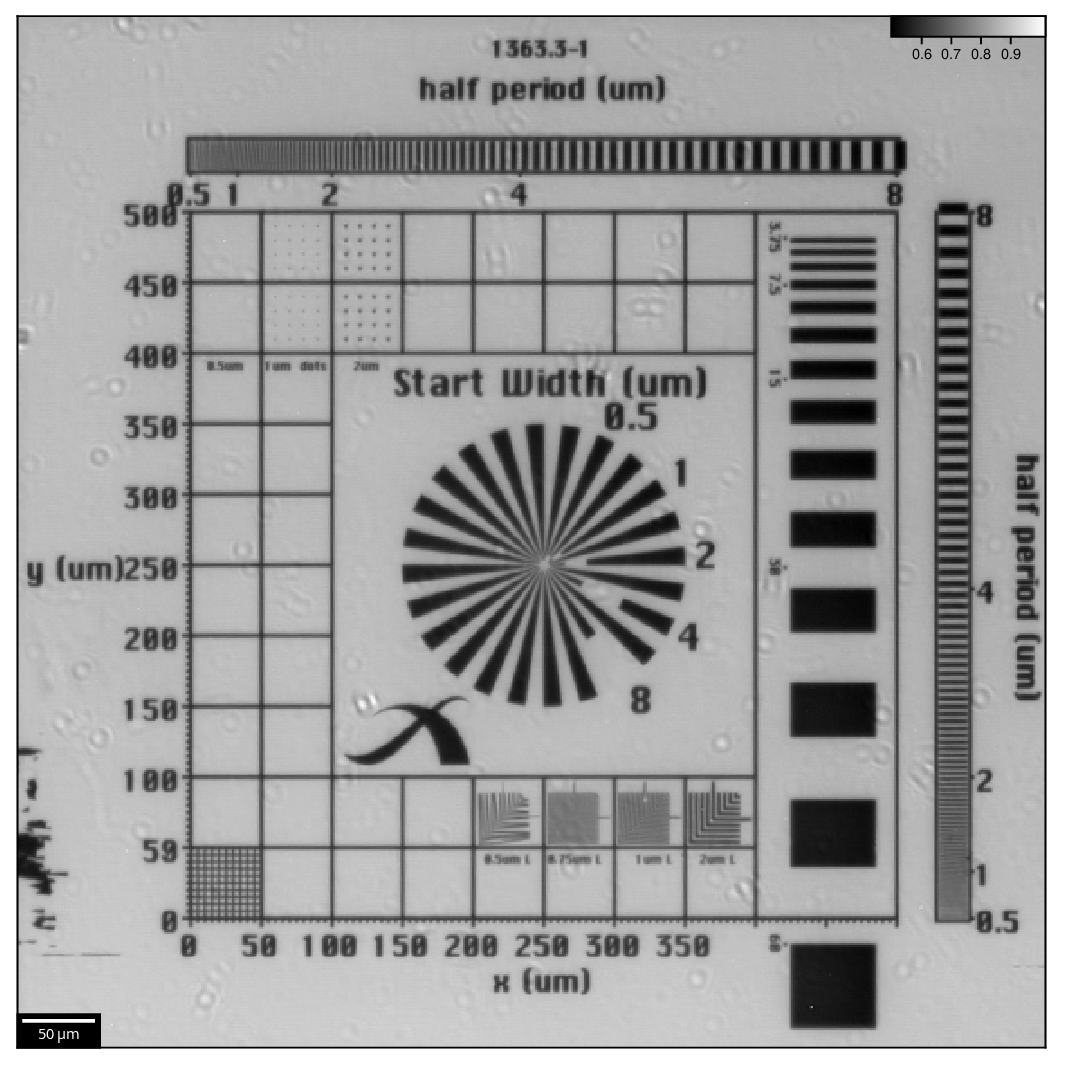}
\renewcommand{\figurename}{Supplementary Fig.}
\caption{Native projection image of the resolution test pattern. The colorbar indicates the normalized intensity.}
\label{fig:supp1}
\end{figure*}

\begin{figure*}[!t]
\centering
\includegraphics[width=440pt]{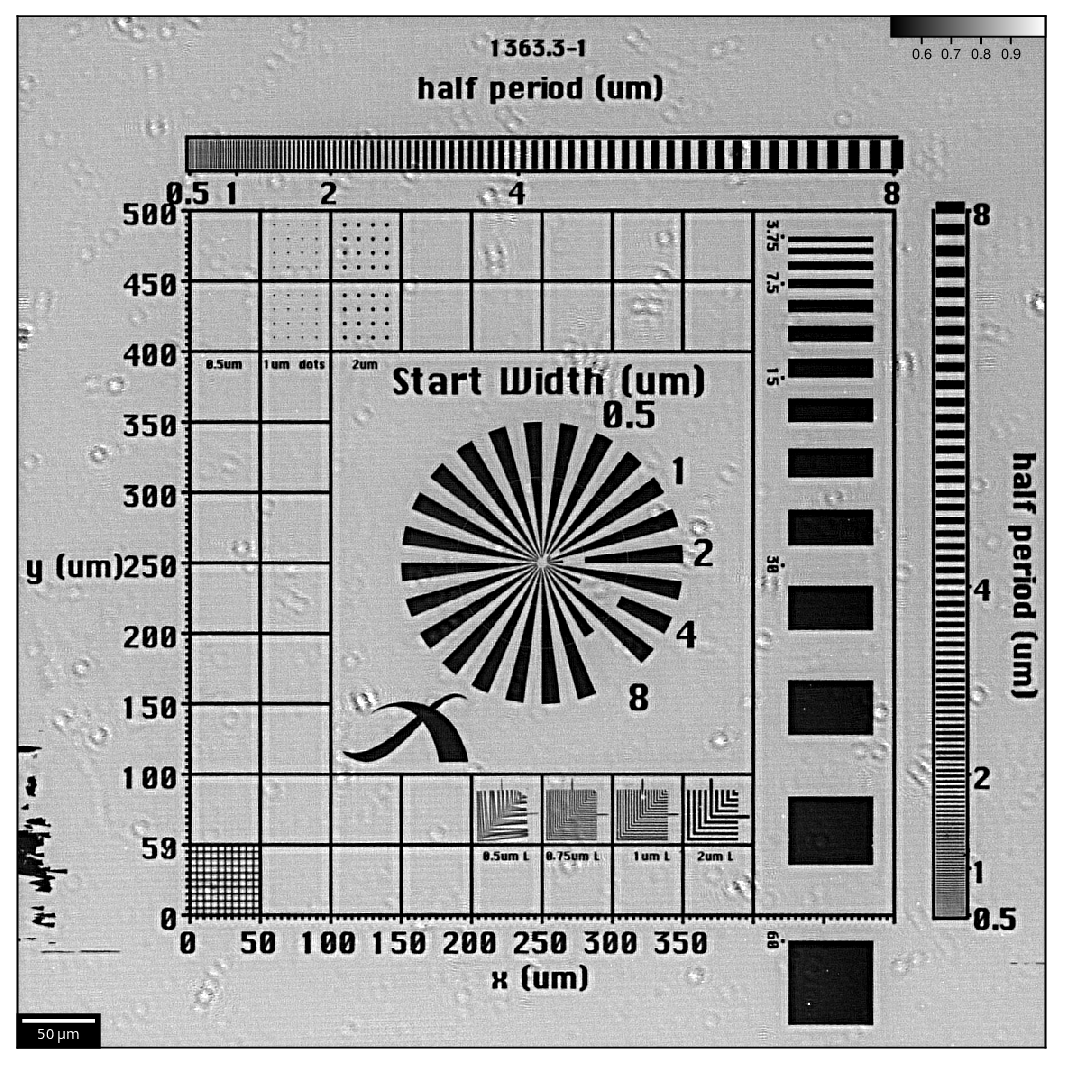}
\renewcommand{\figurename}{Supplementary Fig.}
\caption{Super-resolved projection image of the resolution test pattern. The colorbar indicates the normalized intensity.}
\label{fig:supp2}
\end{figure*}

\begin{figure*}[!t]
\centering
\includegraphics[width=440pt]{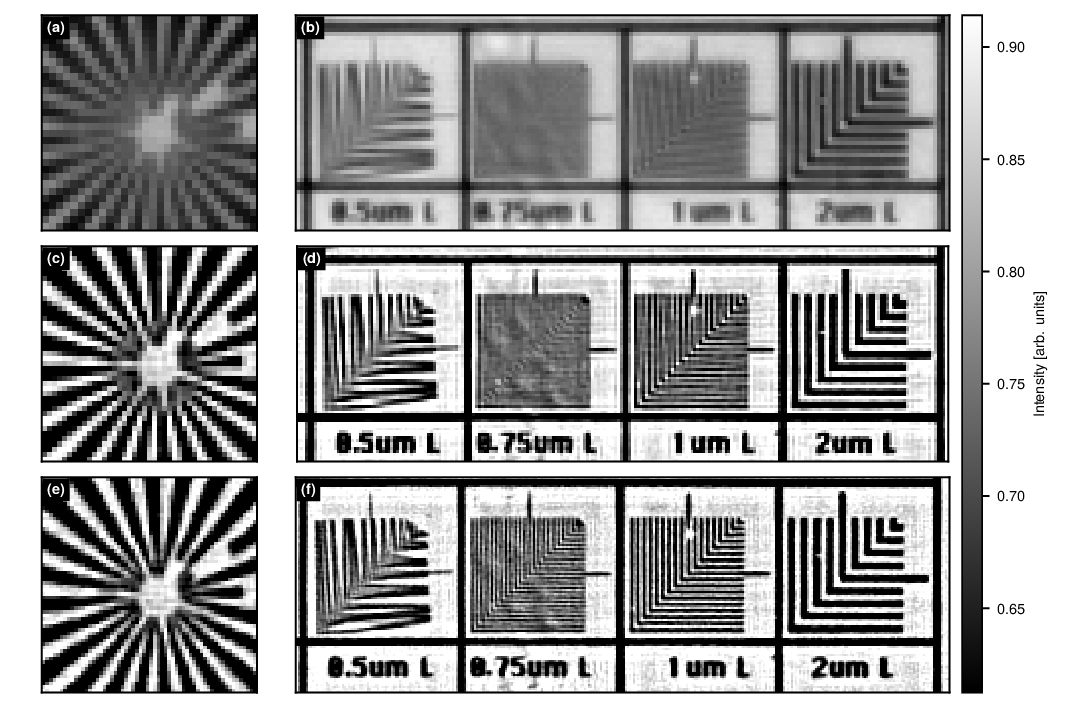}
\renewcommand{\figurename}{Supplementary Fig.}
\caption{Qualitative comparison of resolution gain between first- and second-order reconstruction. (a) Native image of the center region of the Siemens star in the middle of the test pattern. (b) Native image of the L-shaped resolution target.
    (c) and (d) show the first-order super-resolved image of the regions in (a) and (b), respectively. (e) and (f) are the second-order super-resolved images of (a) and (b), respectively.
    While a qualitative assessment shows an improvement in resolution in the first-order reconstruction, as evidenced by the modulation in the \qty{1}{\micro\meter} resolution target, the modulation is higher in the second-order reconstruction.}
\label{fig:supp3}
\end{figure*}

\begin{figure*}[!t]
\centering
\includegraphics[width=440pt]{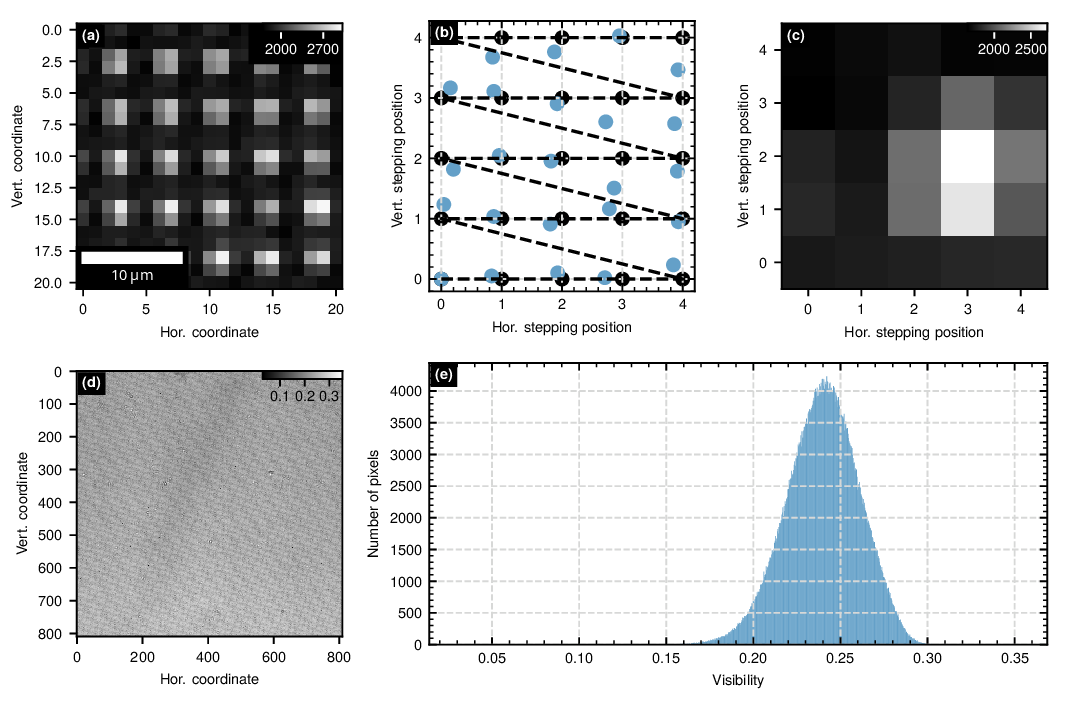}
\renewcommand{\figurename}{Supplementary Fig.}
\caption{Visibility analysis of the utilized grating. (a) Shows an exemplary region of the illumination pattern created by the grating. (b) displays the nominal grating positions relative to those determined by the phase optimization. (c) presents the resulting stepping profile of an arbitrary exemplary pixel. This profile is determined by recording the detected intensity at each pixel as a function of grating position. The colorbars in (a) and (c) indicate the raw detector counts. (d) is the visibility map over the full region in which the test pattern reconstruction was computed. (e) shows the number of pixels per visibility determined from their stepping profile.}
\label{fig:supp4}
\end{figure*}

\begin{figure*}[!t]
\centering
\includegraphics[width=440pt]{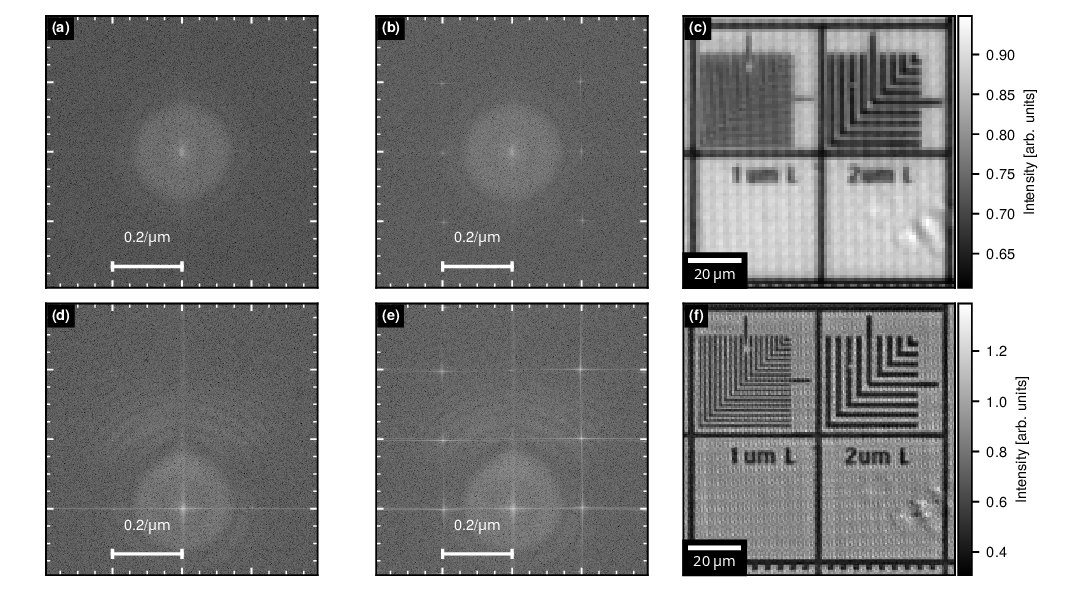}
\renewcommand{\figurename}{Supplementary Fig.}
\caption{Effects of inaccurate phase estimation in the separation matrix. (a) shows the logarithmic absolute value of the separated zero-order component of the reference scan after optimizing the phase values following the approach of Wicker et al.\ \cite{wicker_non-iterative_2013}. (b) shows the same  component, separated under the assumption of equidistant stepping. (c) presents the native projection image computed as the absolute value of the inverse Fourier transform of (b). Inaccurate phase estimation in the separation matrix leads to a residual grating structure. (d) and (e) display the logarithmic absolute value of the first-order components of the reference scan again with (d) and without (e) phase optimization. (f) displays the second-order super-resolved reconstruction under the assumption of equidistant stepping. This inaccurate assumption leads to the visible grating remnants and less resolution gain.}
\label{fig:supp4}
\end{figure*}

\end{document}